\documentclass[5p,times,sort&compress,twocolumn,preprint]{elsarticle}
\usepackage{bm}
\usepackage{amsmath} 
\usepackage{amssymb}
\usepackage{amsthm}
\usepackage{color}
\usepackage{hyperref}
\hypersetup{colorlinks = true}
\hypersetup{breaklinks=true} 
\usepackage{breakurl}

\begin{document}

\begin{frontmatter}

  \title{
    Characteristic time of transition from write error to retention error
    in voltage-controlled magnetoresistive random-access memory
    \tnoteref{grant}
  }

  \author{Hiroko Arai\corref{cor1}}
  \ead{arai-h@aist.go.jp}

  \author{Hiroshi Imamura}
  \ead{h-imamura@aist.go.jp}

  \address{National Institute of Advanced Industrial Science and Technology (AIST),
    Research Center for Emerging Computing Technologies, Tsukuba, Ibaraki, 305-8568, Japan}

  \tnotetext[grant]{
    This work was partly supported by JSPS KAKENHI Grant Number JP19H01108, and JP20K12003.}
  \cortext[cor1]{corresponding author}

  \begin{abstract}
    Voltage controlled magnetoresistive random access memory (VC MRAM) is a promising candidate
    for a future low-power high-density memory.
    The main causes of bit errors in VC MRAM are write error and retention error.
    As the size of the memory cell decreases, the data retention time decreases,
    which causes a transition from the write-error-dominant region to the retention-error-dominant region
    at a certain operating time.
    Here we introduce the characteristic time of the transition
    from the write-error-dominant region to the retention-error-dominant region
    and analyze how the characteristic time depends on the effective anisotropy constant,
    $K_{0}$.
    The characteristic time is approximately expressed as $t_{\rm c} = 2\, w\, \tau$,
    where $w$ is the write error rate,
    and $\tau$ is the relaxation time derived by Kalmkov
      [J. Appl. Phys. 96, (2004) 1138-1145].
    We show that for large $K_{0}$,
    $t_{\rm c}$ increases with increase of $K_{0}$ similar to $\tau$.
    The characteristic time is a key parameter for designing the VC MRAM
    for the variety of applications such as machine learning and artificial intelligence.
  \end{abstract}


\end{frontmatter}



\section{Introduction}
\label{intro}

Keeping information for enough time to ensure the reliability for computing is a key policy of the modern computing systems
based on so-called the von Neumann architecture.
Hierarchical memory system with cache, main memory, and storage
has been successfully realized an effective and reliable computing
\cite{wang_road_2022, sebastian_memory_2020}.
However,
the performance of the von Neumann architecture is limited not only by the performance of the central processing unit (CPU)
but by the data transfer rate between the memory and the CPU through the system bus,
which is known as the von Neumann bottleneck.

In-memory computing is a promising approach to alleviate the von Neumann bottleneck,
where resistive memory devices are used for both processing and memory \cite{ielmini_in-memory_2018,sebastian_memory_2020}.
In-memory  computing  generally  requires  fast,  high-density,  low-power,  scalable  resistive memory  devices,
such  as  resistive random access memory (ReRAM)
\cite{waser_nanoionics-based_2007,chen_reram_2020},
phase-change memory  (PCM)
\cite{raoux_phase_2010, burr_neuromorphic_2017, slesazeck_nanoscale_2019, joshi_accurate_2020},
ferroelectric RAM (FeRAM)
\cite{mikolajick_feram_2001},
and magetoresistive RAM (MRAM)
\cite{yuasa_future_2013, ando_spin-transfer_2014, kent_new_2015, apalkov_magnetoresistive_2016,jain_computing_2018,chang_pxnor-bnn_2019,nozaki_recent_2019,pham_stt-mram_2021}.
A crosspoint  array of these resistive memory devices provides a hardware accelerator for the matrix-vector multiplication (MVM).

Recently much attention has been focused on the in-memory computing
because it provides a powerful MVM tool for machine learning tasks
such as image recognition, object detection, voice recognition, and time-series data analysis
\cite{burr_neuromorphic_2017,lecun_deep_2015,schmidhuber_deep_2015}.
Because the final outputs of these machine learning tasks are provided as a probability and its algorithms are error tolerant
\cite{torres-huitzil_fault_2017,qin_improving_2018},
the accuracy required for the memory is not very high to obtain a practical result.
T. Hirtzlin et al. showed that up to 0.1\% bit error rate can be tolerated
with little impact on recognition performance of a standard binary neural network
\cite{hirtzlin_implementing_2019}.

Voltage-control (VC) MRAM is a promising candidate for the key element of the fast and low-power in-memory computing
because of the fast ($\sim$ 0.1 ns) and low-power ($\sim$ 1 fJ) writing
\cite{maruyama_large_2009,nowak_demonstration_2011,shiota_induction_2012,grezes_ultra-low_2016,kanai_electric-field-induced_2016,yamamoto_improvement_2019,matsumoto_methods_2019, yamamoto_voltage-driven_2020,matsumoto_low-power_2020}.
The VC MRAM uses the voltage control of magnetic anisotropy (VCMA) effect
\cite{maruyama_large_2009,shiota_induction_2012,fahler_electric_2007,nozaki_voltage-induced_2010,miwa_voltage_2017}
to induce the precession of magnetization for writing,
and little Joule heating is generated.

The main causes of bit errors in VC MRAM are write error and retention error.
The write error in VC MRAM is a failure in magnetization switching
due to thermal agitation field during the precession and relaxation.
The write error rate (WER) of the VC MRAM is of the order of $10^{-6}$
\cite{yamamoto_improvement_2019} 
which is low enough to obtain  practical recognition accuracy.
The retention error in VC MRAM is an accidental switching of  magnetization that occurs while retaining the data.
The origin of the retention error is also thermal agitation field.
The retention error rate (RER) of the VC MRAM can be calculated by solving the Fokker-Planck equation
as shown in Ref. \cite{kalmykov_relaxation_2004}.
As the size of the memory cell decreases,
the data retention time decreases,
which causes a transition from the write-error-dominant region to the retention-error-dominant region
at a certain operating time.
For developing the fast and low-power in-memory computing
it is important to know the characteristic time of the transition.

In this paper we define the characteristic time,
$t_{\rm c}$,
of the transition from the write-error-dominant region to the retention-error-dominant region
and analyze the dependence of the characteristic time on the effective anisotropy constant,
$K_{0}$.
We obtained the approximate expression of the characteristic
and show that $t_{\rm c}$ increases with increase of $K_{0}$ for large value of $K_{0}$.

\section{Model and Methods}

\begin{figure}[t]
  \centerline{
    \includegraphics[width=\columnwidth]{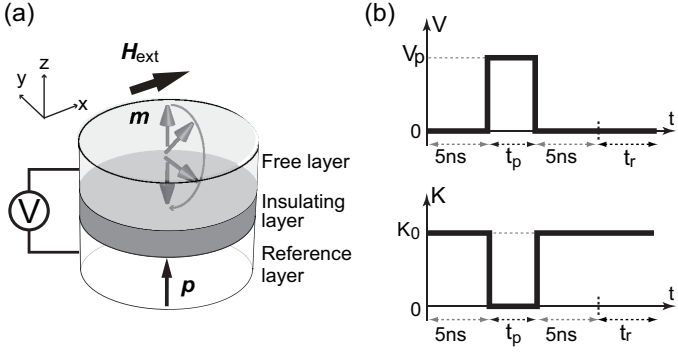}
  }
  \caption{
    (a) Magnetic tunnel junction nanopillar and the definition of Cartesian coordinates
    $(x,y,z)$.
    The non-magnetic insulating layer is sandwiched by the two ferromagnetic layers:
    the free layer and the reference layer.
    A static magnetic field,
    $\bm{H}_{\rm ext}$,
    is applied along the positive $x$ direction.
    Application of the voltage,
    $V$,
    pulse can induce precessional motion of the magnetization
    in the free layer through the VCMA effect.
    (b) Time evolution of the voltage pulse $V$
    and the anisotropy constant $K$.
    Without application of voltage the effective anisotropy constant is $K_{0}$.
    After 5 ns relaxation the voltage pulse with the magnitude of $V_{\rm p}$ is applied
    for the duration of $t_{\rm p}$ to reduce $K$ to zero.
    After the pulse the magnetization is relaxed.
    The WER is calculated using the direction of magnetization at 5 ns after the pulse.
    The contribution from the retention error is estimated by analyzing the error rate
    after the relaxation for the extra relaxation time,
    $t_{\rm r}$.
    \label{fig1}
  }%
\end{figure}

Figure \ref{fig1}(a) schematically shows the magnetic tunnel junction nanopillar
which is the basic element of the VC MRAM.
The nonmagnetic insulating layer is sandwiched by the two ferromagnetic layers
called the free layer and the reference layer.
The size of the MTJ nanopillar is assumed to be so small that the magnetization dynamics can be described
by the macrospin model.
We denote the direction of the magnetization in the free layer by the unit vector
${\bm m}=(m_x,\ m_y,\ m_z)$.
Application of the voltage,
$V$,
pulse reduces the magnetic anisotropy through the VCMA effect
and induces the precession of magnetization around the static external field,
${\bm H}_{\rm ext}$,
applied in the positive $x$ direction.
The magnetization unit vector in the reference layer,
$\bm{p}$,
is fixed to the positive $z$ direction,
i.e. $\bm{p}=(0,0,1)$.
The information stored in the VC MRAM is read as the change of the tunnel resistance
due to the tunnel magnetoresistnace effect
\cite{julliere_tunneling_1975,miyazaki_giant_1995,yuasa_giant_2004,parkin_giant_2004}.

Figure \ref{fig1}(b) shows the shape of the voltage pulse
and the corresponding time evolution of the effective anisotropy constant,
$K$,
in our simulations.
The initial direction of $\bm{m}$ at $t=0$ is set to the equilibrium direction
with $m_{z}>0$,
which minimizes the energy density
\begin{align}
  E=K_{0}(1-m_{z}^{2})-\mu_{0}M_{\rm s} H_{\rm ext} m_{x},
\end{align}
where $K_{0}$ is the effective anisotropy constant of the free layer at $V=0$,
which consists of the crystal magnetic anisotropy and the demagnetizing energy density,
$\mu_{0}$ is the permeability of vacuum,
$M_{\rm s}$ is the saturation magnetization.
For the first 5 ns,
the initial state are thermalized by the thermal agitation field
to obey the Boltzmann distribution at temperature,
$T$.
During the thermalization no voltage is applied,
and the effective anisotropy constant is $K_{0}$.
Then the voltage pulse with the magnitude of $V_{\rm p}$ is applied for the duration of $t_{\rm p}$.
During the pulse the effective anisotropy constant is zero,
and the magnetization precesses around the external magnetic field.
The pulse width,
$t_{\rm p}$,
is set to a half of the precession period to switch the magnetization.
After the pulse the magnetization is relaxed under the condition of $K=K_{0}$.
The WER is calculated using the direction of magnetization at 5 ns
after the end of the pulse.
The contribution from the retention error is estimated by analyzing the error rate
after the relaxation for the extra relaxation time, $t_{\rm r}$.

Temporal evolution of $\bm{m}$ is obtained by solving the Landau-Lifshitz-Gilbert (LLG) equation,
\begin{equation}
  \label{eq:llg}
  \frac{d\bm{m}}{dt}
  = -\gamma \bm{m}\times \bm{H}_{\rm eff}
  +\alpha \bm{m}\times\frac{d\bm{m}}{dt},
\end{equation}
where $\gamma$ is the gyromagnetic constant
and $\alpha$ is the Gilbert damping constant.
The first and second terms on the right hand side represent the torque
resulting from the effective field,
$\bm{H}_{\rm eff}$,
and the damping torque, x
respectively.
The effective field comprises the external field,
anisotropy field,
$\bm{H}_{\rm anis}$,
and thermal agitation field,
$\bm{H}_{\rm therm}$,
as
\begin{equation}
  \bm{H}_{\rm eff}
  = \bm{H}_{\rm ext} + \bm{H}_{\rm anis} + \bm{H}_{\rm therm}.
\end{equation}
The anisotropy field is defined as
\begin{equation}
  \bm{H}_{\rm anis}
  = \frac{2 K(t)}{\mu_{0}M_{\rm s}}m_{z}(t)\, \bm{e}_{z},
\end{equation}
where $\bm{e}_{z}$ is the unit vector in the positive $z$ direction.
The thermal agitation field is determined by the fluctuation-dissipation theorem
\cite{brown_thermal_1963,callen_irreversibility_1951,callen_theorem_1952,callen_statistical_1952,greene_theorem_1952}
and satisfies the following relations
\begin{align}
   & \left\langle H_{\rm therm}^{i}(t) \right\rangle = 0, \\ &
  \left\langle H_{\rm therm}^{i}(t)\, H_{\rm therm}^{j}(t')\ \right\rangle = \xi\,\delta_{i,j}\, \delta(t-t'),
\end{align}
where $\langle\ \rangle$ denotes the statistical average,
the indices $i$, $j$ denote the $x$, $y$, and $z$ components of the thermal agitation field.
$\delta_{i,j}$ represents Kronecker's delta,
and $\delta(t-t')$ represents Dirac's delta function.
The coefficient $\xi$ is given by
\begin{equation}
  \xi
  = \frac{2\alpha k_{\rm B} T}{\gamma\, \mu_{0}\, M_{\rm s}\,\Omega},
\end{equation}
where $k_{\rm B}$ is the Boltzmann constant,
$T$ is temperature,
and $\Omega$ is the volume of the free layer.
The WER is obtained by counting the number of failure in $10^{7}$ trials of writing.
The success or failure of each trial is determined by the sign of $m_{z}$ at $t = t_{\rm p} +10$ ns
  [see Fig. \ref{fig1}(b)].
We also calculate the error rate after $t = t_{\rm p} + 10 $ ns
by  performing the same simulation until $t = t_{\rm p}+t_{\rm r} + 10$ ns.

The RER is the probability of the accidental switching of magnetization by thermal agitation
during the period of data retention.
We employ the theory of RER given by Kalmykov in Ref. \cite{kalmykov_relaxation_2004}, to save the simulation time, and to provide a theoretical analysis.
Introducing the escape ratio,
$\Gamma$,
over the potential barrier
which separate the two different equilibrium directions of the magnetization,
the RER is given by the switching probability defined as
\begin{align}
  \label{eq:pt}
  P(t)
  =\frac{1}{2}\left(
  1-e^{-2\Gamma t}
  \right),
\end{align}
which is the solution of the master equation as
\begin{align}
  \frac{d}{dt}P(t)
  =-\Gamma P(t)+\Gamma (1-P(t)).
\end{align}
The inverse of $2\Gamma$ is called the relaxation time
and is denoted by $\tau$,
i.e., $\tau=1/(2\Gamma)$.

Kalmykov derived the expressions of $\tau$
for three different dissipation regions:
the very low damping (VLD, $\alpha < 0.001$) region,
the intermediate-to-high damping (IHD, $\alpha \gg 1$) region,
and the crossover ($0.01 < \alpha < 1$) region.
For VC MRAM,
the typical value of $\alpha$ is in the crossover region.
The expression of $\tau$ in the crossover region is given by Eq. (18) in Ref. \cite{kalmykov_relaxation_2004} as
\begin{align}
  \tau
  =\tau_{\rm IHD}\frac{A(\alpha S_1+\alpha S_2)}{A(\alpha S_1)A(\alpha S_2)},
  \label{eq:tau}
\end{align}
where
\begin{align}
  A(\alpha S_i)
  =\exp\left[
    \frac{1}{\pi}\int_{0}^{\infty}\, \frac{\ln[1-\exp\{ -\alpha S_i (x^2+1/4) \}]}{x^2+1/4}dx
    \right],
\end{align}
\begin{align}
  \tau_{\rm IHD} \sim \frac{2\tau_{N}\,\pi\sqrt{h}\, e^{\sigma(1-h)^2}}{\sigma\sqrt{1+h} (1-2h+\sqrt{1+4h(1-h)/\alpha^2})},
\end{align}
and
\begin{align}
  S_i
  = & 16\sigma\sqrt{h}\,
  \left[
  1-\frac{13}{6}h+\frac{11}{8}h^2-\frac{3}{16}h^3+\frac{7}{384}h^4 \right.\nonumber \\
    & \left.+\frac{h^5}{256} +O(h^6)
    \right].
\end{align}
Here,
$\tau_{N}=\beta M_{\rm s} (1+\alpha^2)/(2\gamma \alpha)$,
$h=\beta \mu_0 M_{\rm s} H_{\rm ext}/(2\sigma)$,
$\sigma=\beta K$,
and $\beta=\Omega/(k_{\rm B}T)$.
The index of $S_{i}$ is the label of the equilibrium directions,
e.g.,
"1" for $m_{z}>0$
and "2" for $m_{z}<0$.
Since the system we consider has an inversion symmetry with respect to the $x-y$ plane,
i.e.,
$z\to -z$,
Eq. \eqref{eq:tau} can be expressed as
$\tau=\tau_{\rm IHD}A(2\alpha S_1)/A(\alpha S_1)^2$.

The following parameters are assumed.
The saturation magnetization is $M_{\rm s}$ = 0.955 MA/m
\cite{yamamoto_write-error_2019},
and the pulse width is $t_{\rm p}=0.18$ ns.
The magnitude of the external field is $H_{\rm ext}$ = 1 kOe.
The diameter and thickness of the free layer are 40 nm and 1.1 nm,
respectively.
The Gilbert damping constant  $\alpha$ = 0.1,
and temperature is $T=$ 300 K.

\section{Results}

\begin{figure}[t]
  \includegraphics[width=\columnwidth]{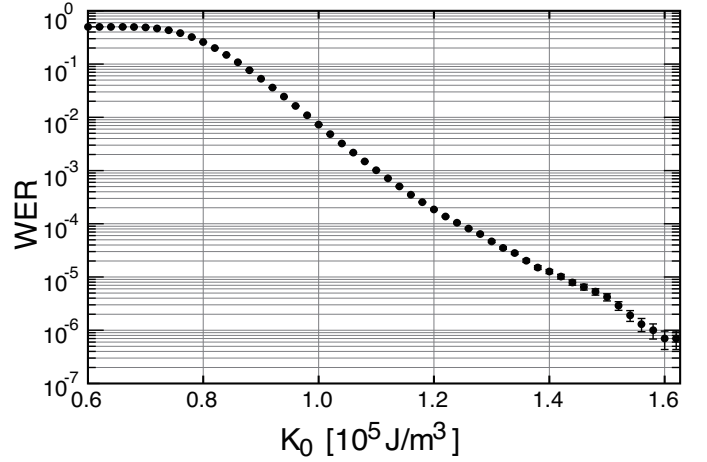}
  \caption{
    WER as a function of anisotropy constant, $K_{0}$.
    The dot represents the mean of the WER,
    and the error bar stands for the standard deviation of the WER.
    Each data points are obtained by averaging the results of $10^7$ trials.
    \label{fig2}
  }
\end{figure}
%
Figure \ref{fig2} shows $K_{0}$ dependence of the WER,
where the dot represents the mean of the WER,
and the error bar stands for the standard deviation of the WER.
Each data points are obtained  by averaging the results of $10^7$ trials.
The WER is about 0.5 for small $K_0$ $(< 0.8)$
because the anisotropy energy is too small to retain the direction of the initial magnetization for 5 ns.
The magnetization is almost equally distributed around the equilibrium directions
with $m_{z} >0$ and $m_{z} <0$ at the beginning of the pulse.
In other words,
the retention time is less than 5 ns.
For $K_0 > 0.8$,
the WER exponentially decreases with increase of $K_{0}$
and reaches below $10^{-6}$
at $K_{0}=1.6\times10^5$ J/m$^{3}$.

%
\begin{figure}[t]
  \includegraphics[width=\columnwidth]{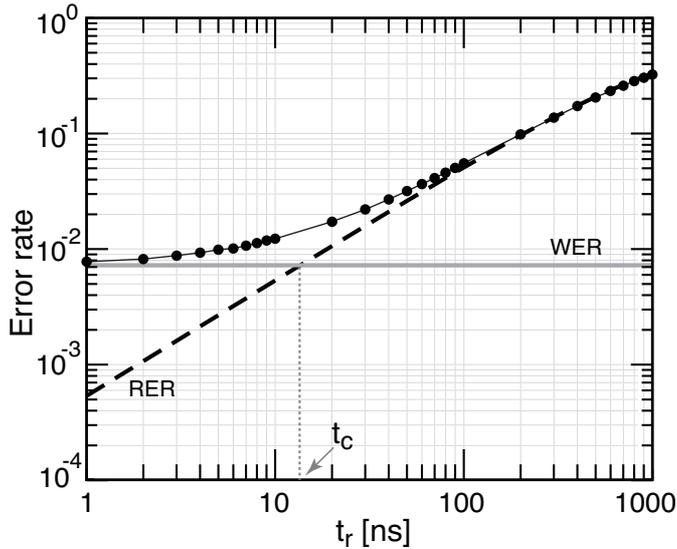}%
  \caption{
    Dependence of the error rate on $t_{\rm r}$
    for $K=1.0\times 10^5$.
    The dots represents the results obtained by numerically solving the LLG equation.
    The thin solid curve connecting the dots is the guide for eyes.
    The Dashed curve represents the RER of Eq. \eqref{eq:pt}.
    The gray horizontal solid line indicates the value of WER:
    $7.3 \times 10^{-3}$.
    The intersection point between the gray horizontal line and the dashed line
    defines the characteristic time for transition
    from the write-error-dominant region to the retention-error-dominant region,
    $t_{\rm c}$.
    The write-error-dominant region is
    $t_{\rm r} < t_{\rm c}$,
    and the retention-error-dominant region is
    $t_{\rm r} > t_{\rm c}$.
    \label{fig3}
  }%
\end{figure}

Figure  \ref{fig3} shows an example of the typical $t_{\rm r}$-dependence of the error rate.
The effective anisotropy constant is assumed to be $K_{0}=1.0\times 10^5$ J/m$^3$.
The dots connected by the solid curve represents the results
obtained by numerically solving the LLG equation.
The gray horizontal solid line indicates the value of the WER:
$7.3 \times 10^{-3}$.
The Dashed curve represents the RER of Eq. \eqref{eq:pt}.
For short $t_{\rm r}$,
the error rate is much larger than the RER
and is dominated by the WER.
For large $t_{\rm r}$,
the error rate converges to the RER,
i.e.,
the RER dominates the error rate.
We introduce the the characteristic time,
$t_{\rm c}$,
for transition from the write-error-dominant region to the retention-error-dominant region
as the intersection point of the dashed curve and the gray horizontal solid line,
as shown in Fig. \ref{fig3}.

Since the RER equals to WER
at $t_{\rm r} = t_{\rm c}$,
we have
\begin{align}
  t_{\rm c}
  =\tau\ln\frac{1}{1-2 w},
  \label{eq:tc}
\end{align}
where $w$ denotes the WER.
Assuming that $w \ll 1$,
Eq. \eqref{eq:tc} is approximated as
\begin{align}
  t_{\rm c} \simeq 2\, w\, \tau.
  \label{eq:tcA}
\end{align}
The $K_{0}$-dependence of $t_{\rm c}$ is determined by the $K_{0}$-dependence of $w$ and $\tau$.
The $K_{0}$-dependence of $w$ is already shown in Fig. \ref{fig2}.
$w$ decreases exponentially with increase of $K_{0}$.

\begin{figure}[t]
  \includegraphics[width=\columnwidth]{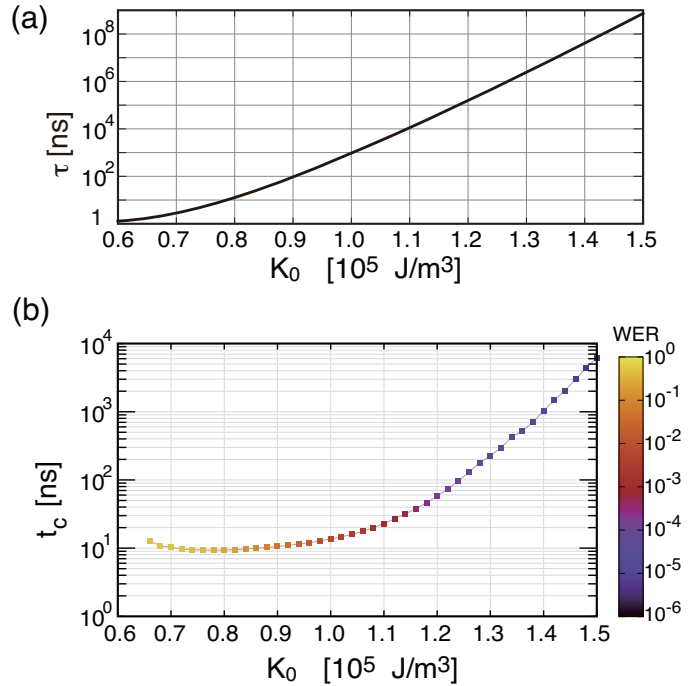}%
  \caption{
    (a) $K_0$-dependence of $\tau$ defined by Eq. \eqref{eq:tau}.
    (b) $K_0$-dependence of characteristic time $t_{\rm c}$.
    Colors indicate the value of WER shown in Fig. \ref{fig2}.
    \label{fig4}
  }%
\end{figure}

The $K_{0}$-dependence of $\tau$ is shown in Fig. \ref{fig4}(a).
On contrary to the $K_{0}$-dependence of $w$,
$\tau$ increases with increase of $K_{0}$.
The $K_{0}$-dependence $t_{\rm c}$ is determined by competition
between the decreasing contribution from $w$ and the increasing contribution from $\tau$.
Figure \ref{fig4}(b) shows the $K_{0}$-dependence of $t_{\rm c}$.
The WER at each point shown in Fig. \ref{fig2} is indicated by color.
Although $t_{\rm c}$ is almost independent of $K_{0}$
for small $K_0\, (< 1.0)$,
it exponentially increases with increase of $K_{0}$
for large $K_0\, (>1.0)$
due to the exponential dependence of $\tau$
on $K_{0}$
shown in Fig. \ref{fig4}(a).

We focused on the influence of $K_0$,
which is the most important parameter of VC MRAM because of its operating principle.
We briefly comment on the influence over $t_{\rm c}$ from another key parameter,
the damping constant,
$\alpha$.
With increase of $\alpha$,
the WER becomes larger.
The present values for the pulse width and the pulse amplitude are the optimal values
that minimize the WER.
The increase of the WER results in the increase of $t_{\rm c}$
(see Fig. \ref{fig3}).

We also briefly comment on the RER of the other MRAM.
The spin-transfer torque (STT) MRAM and spin-orbit torque (SOT) MRAM are principally does not required the application of an external magnetic field,
unlike the VC MRAM.
The analytical formula of the RER for STT/SOT-MRAM has been derived by Brown \cite{brown_thermal_1963}.

\section{Summary}
\label{sec4}
In summary,
we study the characteristic time,
$t_{\rm c}$,
of the transition from write-error-dominant region to retention-error-dominant region
in VC MRAM
by paying special attention to the dependence of $t_{\rm c}$
on the effective anisotropy constant,
$K_{0}$.
We show that
the characteristic time is approximately expressed as
$t_{\rm c} = 2\, w\, \tau$,
where $w$ is the write error rate,
and $\tau$ is the relaxation time
derived by Kalmkov
\cite{kalmykov_relaxation_2004}.
The $K_{0}$-dependence of $t_{\rm c}$ is determined by competition
between the $K_{0}$-dependence of $w$ and $\tau$.
We show that for large $K_0$,
$t_{\rm c}$ increases with increase of $K_0$
similar to $\tau$.
The characteristic time is a key parameter for designing the VC MRAM
for the variety of applications
such as machine learning and artificial intelligence
because the working frequency should be higher than $1/t_{\rm c}$
to ensure the practical recognition accuracy required.

\appendix

\section{Anisotropy constant dependence of half precession period of magnetization}
In VC MRAM, the half precession period is an important parameter.
We briefly discuss the dependence between the half precession period and the anisotropy constant.

For the analysis of the magnetization dynamics during the pulse,
it is convenient to introduce the precession cone angle $\xi$
and the precession angle $\eta$ defined as
$(m_x,\, m_y,\,m_z)=(\cos\xi,\, \sin\xi\cos\eta,\,\sin\xi\sin\eta)$,
and the dimensionless time $\tau=\gamma H_{\rm ext}t/(1+\alpha^2)$.
In terms of $\xi$ and $\eta$,
the LLG equation is expressed as
\begin{align}
  \dot{\xi}  & =-\alpha \sin\xi +\kappa\sin\xi\sin\eta \, (\cos\eta+\alpha\cos\xi\sin\eta)
  \label{xi},                                                                              \\
  \dot{\eta} & =1+\kappa\sin\eta \, (\alpha\cos\eta-\cos\xi\sin\eta),
  \label{eta}
\end{align}
where $\kappa=2K/(\mu_0 M_{\rm s}H_{\rm ext})$.
The initial conditions are $\xi(0)=\arccos(1/\kappa_0)$
and $\eta(0)=\pi/2$,
where $\kappa_0=2K_0/(\mu_0 M_{\rm s}H_{\rm ext})$.
In the case of $K=0$, i.e., $\kappa=0$,
the analytical solutions of $\xi$ and $\eta$ are available,
which represent the damped precession with a spiral trajectory.
The precession angle at the end of pulse,
$\tau=\tau_{\rm p}$,
corresponding to $t=t_{\rm p}$,
is given by $\eta=\pi/2 +\tau_{\rm p}$.
The half precession period is $\tau_{\rm p}=\pi$
($t_{\rm p}=180$ ps in SI unit).

The $K$ dependence of precession period during the pulse in the vicinity of $K=0$
is analyzed by using the perturbation theory.
Expanding the solutions of Eqs. (\ref{xi}) and (\ref{eta}) up to the first order of $\kappa$
and neglecting the term with $\alpha\kappa$,
we obtain
\begin{align}
  \eta(\tau)  \sim  \frac{\pi}{2} + \tau -\frac{\kappa}{4\kappa_0} \, \left\{2\tau + \sin(2\tau) \right\}.
\end{align}
The pulse duration for half of precession period is obtained by solving $\eta(\tau)=3\pi/2$
with the linear approximation of $\sin(2\tau)$ around $\tau=\pi$ as
\begin{align}
  \tau_{\rm p}^{\rm H} =\pi \left( 1+\frac{\kappa}{2\kappa_0}
  \right).
  \label{tauH}
\end{align}
%
%
$\tau_{\rm p}^{\rm H}$ is an increasing function of $\kappa$.

Figure \ref{figA1} shows the value of ($\tau_{\rm p}$, $\kappa$)
satisfying $\eta=3\pi/2$.
The blue thick solid curve shows the approximate result of Eq. (\ref{tauH}). 
The black dashed curve shows the exact result obtained by numerically solving Eqs. (\ref{xi}) and (\ref{eta})
with $\eta(\tau_{\rm p})=3\pi/2$. 
The approximate result of Eq. (\ref{tauH}) 
agrees well with the exact numerical result.

\setcounter{figure}{0}
\begin{figure}[t]
  \includegraphics[width=\columnwidth]{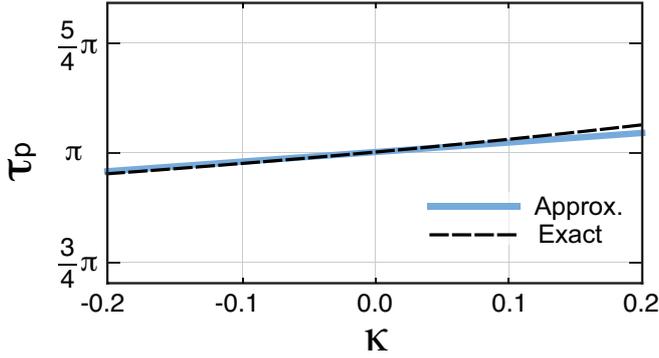}%
  \caption{
    The values of ($\tau_{\rm p}$, $\kappa$) satisfying $\eta=3\pi/2$. 
    The blue solid curve shows the approximate result given by Eq. (\ref{tauH}).
    The black dashed curve represents the exact result obtained by numerically solving Eqs. (\ref{xi}) and (\ref{eta})
    with $\eta(\tau_{\rm p})=3\pi/2$.
    %
    \label{figA1}
  }%
\end{figure}



\begin{thebibliography}{10}
  \expandafter\ifx\csname url\endcsname\relax
    \def\url#1{\texttt{#1}}\fi
  \expandafter\ifx\csname urlprefix\endcsname\relax\def\urlprefix{}\fi
  \expandafter\ifx\csname href\endcsname\relax
    \def\href#1#2{#2} \def\path#1{#1}\fi

  \bibitem{wang_road_2022}
  S.~Wang, X.~Liu, P.~Zhou,
  \href{https://onlinelibrary.wiley.com/doi/10.1002/adma.202106886}{The {Road}
    for {2D} {Semiconductors} in the {Silicon} {Age}}, {Advanced} {Materials} 34
  (2022) 2106886.
  \newline\urlprefix\url{https://doi.org/10.1002/adma.202106886}

  \bibitem{sebastian_memory_2020}
  A.~Sebastian, M.~Le~Gallo, R.~Khaddam-Aljameh, E.~Eleftheriou,
  \href{http://www.nature.com/articles/s41565-020-0655-z}{{Memory} {devices}
    and applications for {in-memory} computing}, Nat. Nanotech. 15 (2020)
  529--544.
  \newline\urlprefix\url{https://doi.org/10.1038/s41565-020-0655-z}

  \bibitem{ielmini_in-memory_2018}
  D.~Ielmini, H.-S.~P. Wong,
  \href{https://doi.org/10.1038/s41928-018-0092-2}{{In-memory} computing with
    resistive switching devices}, Nature Electronics 1 (2018) 333--343.
  \newline\urlprefix\url{https://doi.org/10.1038/s41928-018-0092-2}

  \bibitem{waser_nanoionics-based_2007}
  R.~Waser, M.~Aono, \href{https://doi.org/10.1038/nmat2023}{{Nanoionics}-{based}
  resistive switching memories}, Nat. Mater. 6 (2007) 833--840.
  \newline\urlprefix\url{https://doi.org/10.1038/nmat2023}

  \bibitem{chen_reram_2020}
  Y.~Chen, \href{https://ieeexplore.ieee.org/document/8961211}{{ReRAM}:
  {History}, {Status}, and {Future}}, {IEEE} {Transactions} on {Electron}
    {Devices} 67 (2020) 1420--1433.
  \newline\urlprefix\url{https://doi.org/10.1109/TED.2019.2961505.}

  \bibitem{raoux_phase_2010}
  S.~Raoux, W.~We{\l}nic, D.~Ielmini,
  \href{https://pubs.acs.org/doi/10.1021/cr900040x}{{Phase} {Change}
      {Materials} and {Their} {Application} to {Nonvolatile} {Memories}}, Chem.
  Rev. 110 (2010) 240--267.
  \newline\urlprefix\url{https://doi.org/10.1021/cr900040x}

  \bibitem{burr_neuromorphic_2017}
  G.~W. Burr, R.~M. Shelby, A.~Sebastian, S.~Kim, S.~Kim, S.~Sidler, K.~Virwani,
  M.~Ishii, P.~Narayanan, A.~Fumarola, L.~L. Sanches, I.~Boybat, M.~Le~Gallo,
  K.~Moon, J.~Woo, H.~Hwang, Y.~Leblebici,
  \href{https://www.tandfonline.com/doi/full/10.1080/23746149.2016.1259585}{Neuromorphic
    computing using {non-volatile} memory}, Advances in Physics: X 2 (2017)
  89--124.
  \newline\urlprefix\url{https://doi.org/10.1080/23746149.2016.1259585}

  \bibitem{slesazeck_nanoscale_2019}
  S.~Slesazeck, T.~Mikolajick,
  \href{https://iopscience.iop.org/article/10.1088/1361-6528/ab2084}{Nanoscale
    resistive switching memory devices: a review}, Nanotechnology 30 (2019)
  352003.
  \newline\urlprefix\url{https://doi.org/10.1088/1361-6528/ab2084}

  \bibitem{joshi_accurate_2020}
  V.~Joshi, M.~Le~Gallo, S.~Haefeli, I.~Boybat, S.~R. Nandakumar, C.~Piveteau,
  M.~Dazzi, B.~Rajendran, A.~Sebastian, E.~Eleftheriou,
  \href{https://www.nature.com/articles/s41467-020-16108-9}{Accurate deep
    neural network inference using computational phase-change memory}, Nat.
  Commun. 11 (2020) 2473.
  \newline\urlprefix\url{https://doi.org/10.1038/s41467-020-16108-9}

  \bibitem{mikolajick_feram_2001}
  T.~Mikolajick, C.~Dehm, W.~Hartner, I.~Kasko, M.~Kastner, N.~Nagel, M.~Moert,
  C.~Mazure,
  \href{https://linkinghub.elsevier.com/retrieve/pii/S002627140100049X}{{FeRAM}
    technology for high density applications}, Microelectronics Reliability 41
  (2001) 947--950.
  \newline\urlprefix\url{https://doi.org/10.1016/S0026-2714(01)00049-X}

  \bibitem{yuasa_future_2013}
  S.~Yuasa, A.~Fukushima, K.~Yakushiji, T.~Nozaki, M.~Konoto, H.~Maehara,
  H.~Kubota, T.~Taniguchi, H.~Arai, H.~Imamura, K.~Ando, Y.~Shiota, F.~Bonell,
  Y.~Suzuki, N.~Shimomura, E.~Kitagawa, J.~Ito, S.~Fujita, K.~Abe, K.~Nomura,
  H.~Noguchi, H.~Yoda, \href{http://}{Future prospects of {MRAM} technologies},
  2013 IEEE International Electron Devices Meeting (2013) 3.1.1--3.1.4.
  \newline\urlprefix\url{https://doi.org/10.1109/IEDM.2013.6724549}

  \bibitem{ando_spin-transfer_2014}
  K.~Ando, S.~Fujita, J.~Ito, S.~Yuasa, Y.~Suzuki, Y.~Nakatani, T.~Miyazaki,
  H.~Yoda, \href{http://aip.scitation.org/doi/10.1063/1.4869828}{Spin-transfer
    torque magnetoresistive random-access memory technologies for normally off
    computing (invited)}, J. Appl. Phys. 115 (2014) 172607.
  \newline\urlprefix\url{https://doi.org/10.1063/1.4869828}

  \bibitem{kent_new_2015}
  A.~D. Kent, D.~C. Worledge,
  \href{http://www.nature.com/articles/nnano.2015.24}{A new spin on magnetic
    memories}, Nat. Nanotech. 10 (2015) 187--191.
  \newline\urlprefix\url{https://doi.org/10.1038/nnano.2015.24}

  \bibitem{apalkov_magnetoresistive_2016}
  D.~Apalkov, B.~Dieny, J.~M. Slaughter,
  \href{https://ieeexplore.ieee.org/document/7555318/}{Magnetoresistive
      {Random} {Access} {Memory}}, Proceedings of the IEEE 104 (2016) 1796--1830.
  \newline\urlprefix\url{https://doi.org/10.1109/JPROC.2016.2590142.}

  \bibitem{jain_computing_2018}
  S.~Jain, A.~Ranjan, K.~Roy, A.~Raghunathan, \href{http://}{Computing in
      {Memory} {With} {Spin-Transfer} {Torque} {Magnetic} {RAM}}, IEEE Transactions
  on Very Large Scale Integration (VLSI) Systems 26 (2018) 470--483.
  \newline\urlprefix\url{https://doi.org/10.1109/TVLSI.2017.2776954.}

  \bibitem{chang_pxnor-bnn_2019}
  L.~Chang, X.~Ma, Z.~Wang, Y.~Zhang, Y.~Xie, W.~Zhao,
  \href{http://}{{PXNOR}-{BNN}: {In/With} {Spin}-{Orbit} {Torque} {MRAM}
  {Preset}-{XNOR} {Operation}-{Based} {Binary} {Neural} {Networks}}, IEEE
  Transactions on Very Large Scale Integration (VLSI) Systems 27 (2019)
  2668--2679.
  \newline\urlprefix\url{https://doi.org/10.1109/TVLSI.2019.2926984.}

  \bibitem{nozaki_recent_2019}
  T.~Nozaki, T.~Yamamoto, S.~Miwa, M.~Tsujikawa, M.~Shirai, S.~Yuasa, Y.~Suzuki,
  \href{https://www.mdpi.com/2072-666X/10/5/327}{Recent {Progress} in the
    {Voltage}-{Controlled} {Magnetic} {Anisotropy} {Effect} and the {Challenges}
    {Faced} in {Developing} {Voltage}-{Torque} {MRAM}}, Micromachines 10 (2019)
  327.
  \newline\urlprefix\url{https://doi.org/10.3390/mi10050327}

  \bibitem{pham_stt-mram_2021}
  T.-N. Pham, Q.-K. Trinh, I.-J. Chang, M.~Alioto, \href{http://}{{STT}-{MRAM}
  {Architecture} with {Parallel} {Accumulator} {for} {In-Memory} {Binary}
    {Neural} {Networks}}, IEEE International Symposium on Circuits and Systems
  (ISCAS), (2021) 1--5.
  \newline\urlprefix\url{https://doi.org/10.1109/ISCAS51556.2021.9401695.}

  \bibitem{lecun_deep_2015}
  Y.~LeCun, Y.~Bengio, G.~Hinton,
  \href{http://www.nature.com/articles/nature14539}{Deep learning}, Nature 521
  (2015) 436--444.
  \newline\urlprefix\url{https://doi.org/10.1038/nature14539}

  \bibitem{schmidhuber_deep_2015}
  J.~Schmidhuber,
  \href{https://linkinghub.elsevier.com/retrieve/pii/S0893608014002135}{Deep
  learning in neural networks: {An} overview}, Neural Networks 61 (2015)
  85--117.
  \newline\urlprefix\url{https://doi.org/10.1016/j.neunet.2014.09.003}

  \bibitem{torres-huitzil_fault_2017}
  C.~Torres-Huitzil, B.~Girau,
  \href{http://ieeexplore.ieee.org/document/8013784/}{Fault and {Error}
    {Tolerance} in {Neural} {Networks}: {A} {Review}}, IEEE Access 5 (2017)
  17322--17341.
  \newline\urlprefix\url{https://doi.org/10.1109/ACCESS.2017.2742698}

  \bibitem{qin_improving_2018}
  M.~Qin, C.~Sun, D.~Vucinic, \href{http://}{Improving {Robustness} of {Neural}
      {Networks} against {Bit} {Flipping} {Errors} during {Inference}}, JOIG 6
  (2018) 181--186.
  \newline\urlprefix\url{https://doi.org/10.18178/joig.6.2.181-186}

  \bibitem{hirtzlin_implementing_2019}
  T.~Hirtzlin, B.~Penkovsky, J.-O. Klein, N.~Locatelli, A.~F. Vincent,
  M.~Bocquet, J.-M. Portal, D.~Querlioz,
  \href{https://ieeexplore.ieee.org/document/9073285/}{Implementing {Binarized}
      {Neural} {Networks} with {Magnetoresistive} {RAM} without {Error}
      {Correction}}, 2019 IEEE/ACM International Symposium on Nanoscale
  Architectures (NANOARCH) (2019) 1--5.
  \newline\urlprefix\url{https://doi.org/10.1109/NANOARCH47378.2019.181300}

  \bibitem{maruyama_large_2009}
  T.~Maruyama, Y.~Shiota, T.~Nozaki, K.~Ohta, N.~Toda, M.~Mizuguchi, A.~A.
  Tulapurkar, T.~Shinjo, M.~Shiraishi, S.~Mizukami, Y.~Ando, Y.~Suzuki,
  \href{https://doi.org/10.1038/nnano.2008.406}{Large voltage-induced magnetic
    anisotropy change in a few atomic layers of iron}, Nat. Nanotech. 4 (2009)
  158--161.
  \newline\urlprefix\url{https://doi.org/10.1038/nnano.2008.406}

  \bibitem{nowak_demonstration_2011}
  J.~J. Nowak, R.~P. Robertazzi, J.~Z. Sun, G.~Hu, D.~W. Abraham, P.~L.
  Trouilloud, S.~Brown, M.~C. Gaidis, E.~J. O'Sullivan, W.~J. Gallagher, D.~C.
  Worledge, \href{https://ieeexplore.ieee.org/document/5875962/}{Demonstration
    of ultralow bit error rates for spin-torque magnetic random-access memory
    with perpendicular magnetic anisotropy}, IEEE Magn. Lett. 2 (2011) 3000204.
  \newline\urlprefix\url{https://doi.org/10.1109/LMAG.2011.2155625}

  \bibitem{shiota_induction_2012}
  Y.~Shiota, T.~Nozaki, F.~Bonell, S.~Murakami, T.~Shinjo, Y.~Suzuki,
  \href{https://www.nature.com/articles/nmat3172}{Induction of coherent
    magnetization switching in a few atomic layers of {FeCo} using voltage
    pulses}, Nat. Mater. 11 (2012) 39--43.
  \newline\urlprefix\url{https://doi.org/10.1038/nmat3172}

  \bibitem{grezes_ultra-low_2016}
  C.~Grezes, F.~Ebrahimi, J.~G. Alzate, X.~Cai, J.~A. Katine, J.~Langer,
  B.~Ocker, P.~Khalili~Amiri, K.~L. Wang,
  \href{http://aip.scitation.org/doi/10.1063/1.4939446}{Ultra-low switching
    energy and scaling in electric-field-controlled nanoscale magnetic tunnel
    junctions with high resistance-area product}, Appl. Phys. Lett. 108 (2016)
  012403.
  \newline\urlprefix\url{https://doi.org/10.1063/1.4939446}

  \bibitem{kanai_electric-field-induced_2016}
  S.~Kanai, F.~Matsukura, H.~Ohno,
  \href{http://aip.scitation.org/doi/10.1063/1.4948763}{Electric-field-induced
  magnetization switching in {CoFeB}/{MgO} magnetic tunnel junctions with high
  junction resistance}, Appl. Phys. Lett. 108 (2016) 192406.
  \newline\urlprefix\url{https://doi.org/10.1063/1.4948763}

  \bibitem{yamamoto_improvement_2019}
  T.~Yamamoto, T.~Nozaki, H.~Imamura, Y.~Shiota, S.~Tamaru, K.~Yakushiji,
  H.~Kubota, A.~Fukushima, Y.~Suzuki, S.~Yuasa,
  \href{https://iopscience.iop.org/article/10.1088/1361-6463/ab03c2}{Improvement
    of write error rate in voltage-driven magnetization switching}, J. Phys. D:
  Appl. Phys. 52 (2019) 164001.
  \newline\urlprefix\url{https://doi.org/10.1088/1361-6463/ab03c2}

  \bibitem{matsumoto_methods_2019}
  R.~Matsumoto, H.~Imamura,
  \href{http://aip.scitation.org/doi/10.1063/1.5128154}{Methods for reducing
    write error rate in voltage-induced switching having prolonged tolerance of
    voltage-pulse duration}, AIP Advances 9 (2019) 125123.
  \newline\urlprefix\url{https://doi.org/10.1063/1.5128154}

  \bibitem{yamamoto_voltage-driven_2020}
  T.~Yamamoto, T.~Nozaki, H.~Imamura, S.~Tamaru, K.~Yakushiji, H.~Kubota,
  A.~Fukushima, Y.~Suzuki, S.~Yuasa,
  \href{https://link.aps.org/doi/10.1103/PhysRevApplied.13.014045}{Voltage-{Driven}
  {Magnetization} {Switching} {Using} {Inverse}-{Bias} {Schemes}}, Phys. Rev.
  Applied 13 (2020) 014045.
  \newline\urlprefix\url{https://doi.org/10.1103/PhysRevApplied.13.014045}

  \bibitem{matsumoto_low-power_2020}
  R.~Matsumoto, H.~Imamura,
  \href{https://link.aps.org/doi/10.1103/PhysRevApplied.14.021003}{Low-{Power}
  {Switching} of {Magnetization} {Using} {Enhanced} {Magnetic} {Anisotropy}
  with {Application} of a {Short} {Voltage} {Pulse}}, Phys. Rev. Applied 14
  (2020) 021003.
  \newline\urlprefix\url{https://doi.org/10.1103/PhysRevApplied.14.021003}

  \bibitem{fahler_electric_2007}
  M.~Weisheit, S.~F\"{a}hler, A.~Marty, Y.~Souche, C.~Poinsignon, D.~Givord,
  \href{https://www.science.org/doi/abs/10.1126/science.1136629}{Electric
    {Field}-{Induced} {Modification} of {Magnetism} in {Thin}-{Film}
  {Ferromagnets}}, Science 315 (2007) 349--351.
  \newline\urlprefix\url{https://doi.org/10.1126/science.1136629}

  \bibitem{nozaki_voltage-induced_2010}
  T.~Nozaki, Y.~Shiota, M.~Shiraishi, T.~Shinjo, Y.~Suzuki,
  \href{http://aip.scitation.org/doi/10.1063/1.3279157}{Voltage-induced
    perpendicular magnetic anisotropy change in magnetic tunnel junctions}, Appl.
  Phys. Lett. 96 (2010) 022506.
  \newline\urlprefix\url{https://doi.org/10.1063/1.3279157}

  \bibitem{miwa_voltage_2017}
  S.~Miwa, M.~Suzuki, M.~Tsujikawa, K.~Matsuda, T.~Nozaki, K.~Tanaka,
  T.~Tsukahara, K.~Nawaoka, M.~Goto, Y.~Kotani, T.~Ohkubo, F.~Bonell,
  E.~Tamura, K.~Hono, T.~Nakamura, M.~Shirai, S.~Yuasa, Y.~Suzuki,
  \href{https://www.nature.com/articles/ncomms15848}{Voltage controlled
    interfacial magnetism through platinum orbits}, Nat. Commun. 8 (2017) 15848.
  \newline\urlprefix\url{https://doi.org/10.1038/ncomms15848}

  \bibitem{yamamoto_write-error_2019}
  T.~Yamamoto, T.~Nozaki, H.~Imamura, Y.~Shiota, T.~Ikeura, S.~Tamaru,
  K.~Yakushiji, H.~Kubota, A.~Fukushima, Y.~Suzuki, S.~Yuasa,
  \href{https://link.aps.org/doi/10.1103/PhysRevApplied.11.014013}{Write-{Error}
  {Reduction} of {Voltage}-{Torque}-{Driven} {Magnetization} {Switching} by a
    {Controlled} {Voltage} {Pulse}}, Phys. Rev. Applied 11 (2019) 014013.
  \newline\urlprefix\url{https://doi.org/10.1103/PhysRevApplied.11.014013}

  \bibitem{kalmykov_relaxation_2004}
  Y.~P. Kalmykov, \href{http://aip.scitation.org/doi/10.1063/1.1760839}{The
    relaxation time of the magnetization of uniaxial single-domain ferromagnetic
    particles in the presence of a uniform magnetic field}, J. Appl. Phys. 96
  (2004) 1138--1145.
  \newline\urlprefix\url{https://doi.org/10.1063/1.1760839}

  \bibitem{julliere_tunneling_1975}
  M.~Julliere,
  \href{https://www.sciencedirect.com/science/article/pii/0375960175901747}{Tunneling
    between ferromagnetic films}, Phys. Lett. A 54 (1975) 225--226.
  \newline\urlprefix\url{https://doi.org/10.1016/0375-9601(75)90174-7}

  \bibitem{miyazaki_giant_1995}
  T.~Miyazaki, N.~Tezuka, \href{https://}{Giant magnetic tunneling effect in
    {Fe}/{Al$_2$O$_3$}/{Fe} junction}, J. Magn. Magn. Mat. 139 (1995) L231--L234.
  \newline\urlprefix\url{https://doi.org/10.1016/0304-8853(95)90001-2}

  \bibitem{yuasa_giant_2004}
  S.~Yuasa, T.~Nagahama, A.~Fukushima, Y.~Suzuki, K.~Ando,
  \href{https://www.nature.com/articles/nmat1257}{Giant room-temperature
  magnetoresistance in single-crystal {Fe}/{MgO}/{Fe} magnetic tunnel
  junctions}, Nat. Mater. 3 (2004) 868--871.
  \newline\urlprefix\url{https://doi.org/10.1038/nmat1257}

  \bibitem{parkin_giant_2004}
  S.~S.~P. Parkin, C.~Kaiser, A.~Panchula, P.~M. Rice, B.~Hughes, M.~Samant,
  S.-H. Yang, \href{https://www.nature.com/articles/nmat1256}{Giant tunnelling
    magnetoresistance at room temperature with {MgO} (100) tunnel barriers}, Nat.
  Mater. 3 (2004) 862--867.
  \newline\urlprefix\url{https://doi.org/10.1038/nmat1256}

  \bibitem{brown_thermal_1963}
  W.~F. Brown, \href{https://link.aps.org/doi/10.1103/PhysRev.130.1677}{Thermal
    {Fluctuations} of a {Single}-{Domain} {Particle}}, Phys. Rev. 130 (1963)
  1677--1686.
  \newline\urlprefix\url{https://doi.org/10.1103/PhysRev.130.1677}

  \bibitem{callen_irreversibility_1951}
  H.~B. Callen, T.~A. Welton,
  \href{https://link.aps.org/doi/10.1103/PhysRev.83.34}{Irreversibility and
      {Generalized} {Noise}}, Phys. Rev. 83 (1951) 34--40.
  \newline\urlprefix\url{https://doi.org/10.1103/PhysRev.83.34}

  \bibitem{callen_theorem_1952}
  H.~B. Callen, R.~F. Greene,
  \href{https://link.aps.org/doi/10.1103/PhysRev.86.702}{On a {Theorem} of
      {Irreversible} {Thermodynamics}}, Phys. Rev. 86 (1952) 702--710.
  \newline\urlprefix\url{https://doi.org/10.1103/PhysRev.86.702}

  \bibitem{callen_statistical_1952}
  H.~B. Callen, M.~L. Barasch, J.~L. Jackson,
  \href{https://link.aps.org/doi/10.1103/PhysRev.88.1382}{Statistical
      {Mechanics} of {Irreversibility}}, Phys. Rev. 88 (1952) 1382--1386.
  \newline\urlprefix\url{https://doi.org/10.1103/PhysRev.88.1382}

  \bibitem{greene_theorem_1952}
  R.~F. Greene, H.~B. Callen,
  \href{https://link.aps.org/doi/10.1103/PhysRev.88.1387}{On a {Theorem} of
    {Irreversible} {Thermodynamics}. {II}}, Phys. Rev. 88 (1952) 1387--1391.
  \newline\urlprefix\url{https://doi.org/10.1103/PhysRev.88.1387}

\end{thebibliography}

\end{document}